# Hardware implementation of Bayesian network building blocks with stochastic spintronic devices

P. Debashis[1,2,†], V. Ostwal[1,2,†], R. Faria[1], S. Datta[1], J. Appenzeller[1,2] and Z. Chen[1,2]
[1]School of Electrical and Computer Engineering, [2]Birck Nanotechnology Center
Purdue University, West Lafayette, IN 47907, USA, Email: zhchen@purdue.edu
[†]authors contributed equally

*Abstract*—Bayesian networks are powerful statistical models to understand causal relationships in real-world probabilistic problems such as diagnosis, forecasting, computer vision, etc. For systems that involve complex causal dependencies among many variables, the complexity of the associated Bayesian networks become computationally intractable. As a result, direct hardware implementation of these networks is one promising approach to reducing power consumption and execution time. However, the few hardware implementations of Bayesian networks presented in literature rely on deterministic CMOS devices that are not efficient in representing the inherently stochastic variables in a Bayesian network. This work presents an experimental demonstration of a Bayesian network building block implemented with naturally stochastic spintronic devices. These devices are based on nanomagnets with perpendicular magnetic anisotropy, initialized to their hard axes by the spin orbit torque from a heavy metal under-layer utilizing the giant spin Hall effect, enabling stochastic behavior. We construct an electrically interconnected network of two stochastic devices and manipulate the correlations between their states by changing connection weights and biases. By mapping given conditional probability tables to the circuit hardware, we demonstrate that any two node Bayesian networks can be implemented by our stochastic network. We then present the stochastic simulation of an example case of a four node Bayesian network using our proposed device, with parameters taken from the experiment. We view this work as a first step towards the large scale hardware implementation of Bayesian networks.

## I. Introduction

Bayesian networks (BNs) are directed graphical models that are used to represent the causal dependencies among stochastic variables[1]. In a BN, each node represents a stochastic variable, whose probability of occurrence is determined by the states of its parent nodes. The dependence between a set of such nodes is given by a conditional probability table (CPT). BNs are traditionally implemented in software aiming at applications in areas such as forecasting, diagnosis, and computer vision[2]. However, as the complexity of the BNs grows, i.e., as the number of parent nodes affecting the probability of a particular child node becomes large, both the assessment of that child node probability, and the inference about the possible cause becomes impractical[3]. Specifically, as the network size grows, the number of terms in the calculation of the joint probability using probability chain rule increases rapidly[3].

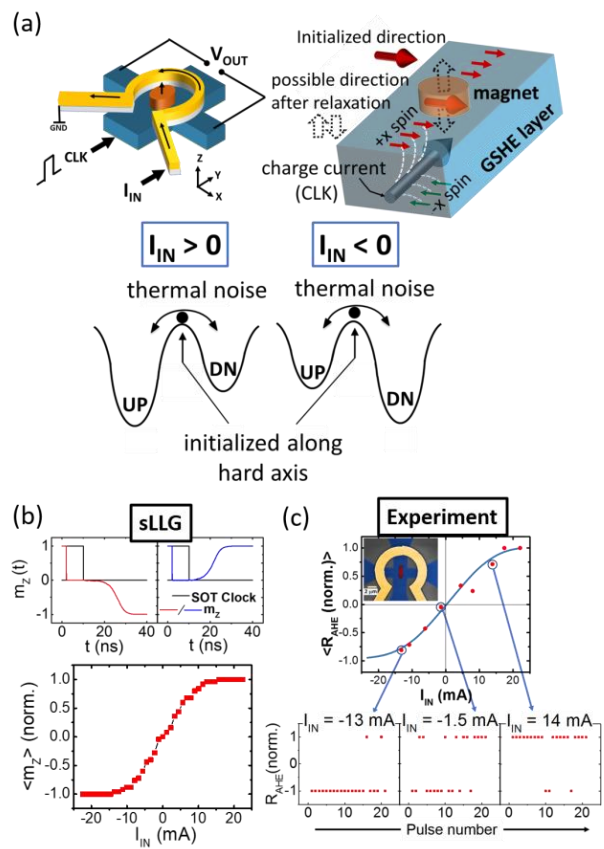

Fig. 1. Hardware building block of Bayesian Networks. (a) Schematic of the probabilistic device and illustration of the hard axis initialization by spin orbit torque. (b) Stochastic LLG simulation of 500 ensembles, showing tunable random behavior of the device. The two top panels show representative cases where the magnetization relaxes to the "up" and "down" direction after being released from the hard axis. (c) Experimental measurements on the device showing stochastic behavior with tunability using a charge current through an isolated Oersted ring. The bottom panels show the stochastic outputs, whose averages show the sigmoidal behavior as a function of the input current.

Direct representation of Bayesian networks in hardware has been proposed as an alternative way to perform the two above mentioned tasks, i.e., probability assessment and inference. In this case, each "node" in a Bayesian network is represented by a stochastic device, having a distinct probability of being in one of two possible states. This probability is controlled by the input it receives dependent on the states of its parent nodes, through the weights of the connections between them. The CPT is encoded in the weights of these connections. By representing a BN with a hardware network of this kind, the required probability of a particular event is readily obtained by sampling the output of the corresponding stochastic device. Moreover, inference about the possible cause of a particular event can be evaluated by observing the joint distribution of the two stochastic devices corresponding to the "event" node and the particular "cause" node of interest.

Several hardware implementations of BNs have been proposed based on CMOS hardware. For example, Zermani et al.[4] demonstrated FPGA based BN implementation utilizing suitable architectural design and memory allocation schemes. Cai et al.[5] demonstrated another FPGA based architectural design along with a suitable pseudo random number generator. Manisinghka et al.[6] implemented a BN in digital circuits using a novel abstraction. Chakrapani et al.[7] and Weijia et al.[8] proposed a probabilistic CMOS hardware for BN implementation, however there has not been an experimental demonstration in literature to our knowledge. Nevertheless, there is an interest in a compact implementation of the stochastic nodes of a BN and their conditional relations.

In this work, we present an experimental demonstration of a spintronics based compact hardware implementation of BNs. The stochastic elements are implemented naturally by a compact device consisting of a perpendicular nanomagnet. The CPTs are translated directly to the connection weights, implemented by resistive connections between such devices.

Unstable nanomagnet based spintronic devices have recently attracted much research interest for probabilistic spin logic (PSL)[9,10,19–21,11–18] and are given the name "p-bit", which is the short form of "probabilistic bit". It has been proposed that inherently unstable nanomagnet can be a natural implementation of the stochastic variable in a BN[9,18,19,22]. We first present a p-bit implementation using a stochastic spintronic device that has isolated input and output to allow for interconnection in circuitry. The output of such a device is a tunable random number, whose mean is controlled by an electrical input. Then, we build an electrically connected network of two such devices and study the correlation of their outputs for different connections and biases. We show that any CPT can be implemented by changing the connections and biases of this circuit, thus representing a hardware BN building block. Finally, using parameters taken from the experiment, we perform a stochastic Landau Lifshitz Gilbert (sLLG) simulation of a four node BN and compare the results of the forecast with those expected from calculating joint probability distributions.

## II. EXPERIMENTAL RESULTS AND ANALYSIS

### A. Hard axis initialized PMA magnet as p-bit

In our experiment, the stochastic device is based on a hard axis initialized magnet with perpendicular magnetic anisotropy (PMA), whose output probability is controlled by the magnetic field produced by a charge current passing through an isolated metal ring[14,15,17]. The top left of Fig. 1 (a) shows the schematic of our device. It consists of a nanomagnet island with perpendicular magnetic anisotropy (PMA) shown in orange, on top of a heavy metal (Ta) Hall bar, shown in blue. It is well understood that the magnetization of a PMA magnet can be deterministically switched by the Spin Obit Torque (SOT) of a heavy metal under-layer in the presence of a symmetry breaking in-plane magnetic field[23,24]. However, when the spin current density is large enough, and when this field is absent, the magnetization gets pinned in the direction of the spin polarization, i.e. the magnets hard axis. Once the spin current is removed, ambient thermal noise makes the magnetization relax to either "up" or "down" with equal probability due to the symmetric energy landscape for these two states[14,15,25] as depicted by the cartoon in the top right of Fig. 1 (a). The magnetization state is read out by the anomalous Hall effect (AHE), where the transverse $V_{OUT}$ is +ve for a magnetization in the "up" direction and -ve for "down". The probability of relaxing back to the "up" or "down" direction can be controlled by applying a small out-of-plane magnetic field that lifts the degeneracy of the energy landscape. A positive field in the z-direction lowers the energy of the "up" state and raises that of the "down" state, thus making the "up" state more favorable. A negative z-directed field does the exact opposite. This is depicted in the energy landscape diagrams shown in the bottom panel of Fig. 1 (a). This z-directed field is provided by a ring-shaped electrode called the "Oersted ring" henceforth, shown in yellow in the device schematic. A current "$I_{IN}$" passing through the Oersted ring of radius "r" produces a magnetic field given by $B = \mu_0 * I_{IN}/2r$.

Fig. 1 (b) shows the sLLG simulation of such a device. The top panels show the magnetization dynamics during the pulsing of the device. The current pulse through the GSHE layer is shown in black color in both the panels. The z-component of magnetization ($m_Z$) is shown in blue and red. It can be seen that $m_Z$ goes to zero while the current pulse is ON. After the pulse is removed, $m_Z$ relaxes to -1 in the first case and it relaxes to +1 in the second, nominally identical case, highlighting the stochastic nature of the process. The time scale of this relaxation is governed by the material parameters of the nanomagnet such as $M_S$, $H_K$ and damping. The bottom panel of Fig. 1 (b) shows the average of the magnetization (after the dynamics have settled) in the z-direction (perpendicular easy axis) as a function of the input current, resembling a sigmoidal activation function.

For experimental implementation, starting with a stack of Ta(5nm)/CoFeB(1nm)/MgO(2nm)/Ta(1nm) thin film, a Hall bar device with a PMA magnetic island located at the center is fabricated by means of successive e-beam lithography and Ar ion milling steps. To generate the out-of-plane field for tunability, the "Oersted ring" is fabricated on top and electrically isolated from the Hall bar by a dielectric layer. A false colored SEM image of the fabricated device is shown in the inset of Fig. 1(c).

For the operation of the device, a Keithley 6221 current source is used to provide a current pulse of duration 100 μs through the Ta Hall bar. This current pulse experimentally implements the required hard axis biasing scheme as shown in the sLLG simulation of Fig. 1 (b). Although the magnet can respond to much faster pulses, as shown in Fig. 1(b), we chose to use 100 μs to be safely within the delay times of the measurement circuit. After the pulsing event, the state of the magnetization is read by a lock-in scheme, with a sinusoidal current provided by the same Keithley current source and an SRS830 lock-in amplifier. The device is pulsed repeatedly, and the state of the magnetization is read after each individual pulse. Fig. 1 (c) shows the average magnetization as a function of the input current "$I_{IN}$". Each data point is obtained by averaging 25 pulsing events, as shown for three representative cases in the bottom panels. These measurements clearly demonstrate the successful implementation of a device with an electrical input and output, which behaves stochastically for individual events, but produces a sigmoidal curve for the average output. This is the desired characteristic for many probabilistic spin logic applications including hardware BNs.

### B. Implementing a two node Bayesian network in hardware

Next, we show how the stochastic devices described in the previous section can be used to implement a two node Bayesian network in hardware. The essential characteristic of a BN is captured in the CPT. Fig. 2 (a) shows the example of a two-node network, with the first or the parent node ($m_1$) representing the packaging material for blocks of cheese in a dairy farm, and the second node ($m_2$) representing the probability of finding a stale cheese block. The values "a" and "b" in the CPT represent the probability of a cheese block being stale if the packaging material is of low quality ($m_1 = 0$) vs. high quality ($m_1=1$). Since the packaging material positively affects the shelf life, in this case, a > b. If instead of packaging material, $m_1$ represents the print design on the package, then the shelf life is not affected by it, and hence, a=b in this case. Similarly, if some other variable, that negatively affects the shelf life is represented by $m_1$, then the CPT would have a < b. Now, for the first case, if the cheese was stored in a cold and dry storage, then the shelf life is increased, irrespective of the packaging material quality. This corresponds to adding a positive value to both "a" and "b" in the CPT. Hence, the variables in the CPT can span the entire space between 0 and 1 independently, depending on the problem being modeled.

We first demonstrate that the CPT between the two probabilistic random variables in our example can be implemented by design of proper electrical connections between two of our stochastic devices (of the type shown in Fig. 1). Then, by testing the circuit with designed parameters, we show that the probability of the output device ($m_2$) follows the probability of finding a stale cheese block, obtained from calculating the joint probability distribution. We also show that the inference about the potential cause of stale cheese that is evaluated by Bayes theorem is well matched to the directly observed values from the joint distribution of the device outputs. The results are also verified by stochastic LLG simulations with magnet parameters ($M_S$, $H_K$ and volume) taken to match the sigmoidal activation function obtained from the experiment.

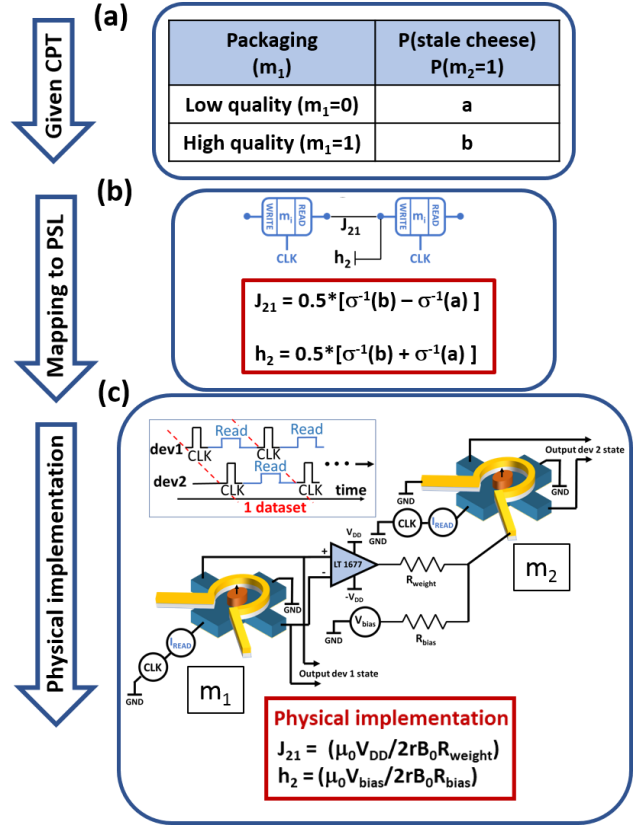

Fig. 2. Hardware design of a two-node network. (a) The given conditional probability table (CPT) representing the causal dependency of two probabilistic variables, i.e., the quality of packaging and state of cheese (b) PSL model of the two node BN with the CPT parameters translated to PSL parameters (c) Circuit schematic of two connected devices to implement two coupled Bayesian nodes. Inset on the top left shows the timing diagram of various operations performed on device 1 and 2.

Fig. 2 (a) shows the given CPT that represents the relation between the stochastic variables $m_1$ and $m_2$. This CPT is translated into the parameters $J_{21}$ and $h_2$ of the PSL model as shown in Fig. 2 (b). This translation can be obtained from the analysis below:

The total input, $I_2$, received by the second device is given by:

$$I_2 = J_{21}*m_1 + h_2 \qquad (1)$$

where $J_{21}$ corresponds to the connection from the first to the second device, $m_1$ corresponds to the state of the first device and $h_2$ corresponds to the constant bias given to the second device. As eq. 1 represents the physical input to node 2 (which is current in our hardware design), $m_1$ has to enter as a bipolar value (+1 for 'UP' state and -1 for 'DN' state).

The average state of the second device is given by:

$$\langle m_2 \rangle = \sigma(I_2) = \sigma(J_{21}*m_1 + h_2) \quad (2)$$

where $\sigma$ represents the sigmoidal activation function for device 2. The conditional dependencies can be directly seen from this expression. The probability of $m_2$ being high given $m_1$ high is obtained by evaluating $\langle m_2 \rangle$ from equation (2) by setting $m_1 = 1$. Since this probability should match the value specified in the given CPT, we obtain:

$$b = \sigma(J_{21} + h_2) \quad (3)$$

Similarly, 'a' can be obtained by setting $m_1 = -1$ (as bipolar entry corresponding to $m_1$ being 'DN' is -1 instead of 0) in equation (2)

$$a = \sigma(-J_{21} + h_2) \quad (4)$$

From equations (3) and (4), we obtain the values of the PSL parameters $J_{21}$ and $h_2$, from the given CPT table as follows:

$$J_{21} = 0.5*[\sigma^{-1}(b) - \sigma^{-1}(a)] \quad (5)$$
$$h_2 = 0.5*[\sigma^{-1}(b) + \sigma^{-1}(a)] \quad (6)$$

The parameters $J_{21}$ and $h_2$ are then used to design the hardware connection strengths and biases to two stochastic devices, as will be discussed in the following paragraphs.

Fig. 2 (c) shows the schematic of our circuit. The output voltage from the first device is amplified by a LT1677 low noise, rail-to-rail precision Op Amp operating in an open loop configuration. The output level of the Op Amp is determined by its +/- $V_{DD}$ supply voltages. This output is then connected to the Oersted ring of the second device through a weight resistor "$R_{weight}$" that determines how much current passes through it, and hence controls the output probability of the second device, corresponding to the $J_{21}$ term in a BN. Additionally, a voltage source "$V_{bias}$" is connected to the input of the second device through a resistor "$R_{bias}$" to mimic the fixed bias ($h_2$) in a BN. The values of the circuit parameters $V_{DD}$, $V_{bias}$, $R_{weight}$ and $R_{bias}$ are obtained from the required $J_{21}$ and $h_2$ by the following design analysis:

In our circuit as shown in Fig. 2(c), $J_{21}$ is the magnetic field produced by the Oersted ring of device 2, normalized with the field required to saturate its magnetization in the "up" or "down" state, denoted by $B_0$. This is given by:

$$J_{21} = \pm \mu_0 V_{DD}/2rB_0 R_{weight} \quad (7)$$

where r is the radius of the Oersted ring, $\mu_0$ is the permeability of vacuum and the $\pm$ sign depends on the connection polarity. Similarly, $h_2$ is the additional magnetic field produced by the constant bias $V_{bias}$, normalized to $B_0$.

$$h_2 = (\mu_0 V_{bias}/2rB_0 R_{bias}) \quad (8)$$

Note that $h_2$ contributions due to the remnant magnetic field in the measurement setup have been subtracted out in this analysis for brevity. This additional $h_2$ contribution is just added to the calculated $h_2$ in equation 8.

Next, we show that the same circuit can capture any given CPT, by changing the $R_{weight}$ and $R_{bias}$. In the circuit shown in Fig. 2(c), the total input received by device 2 is given by:

$$I_2 = \pm(\mu_0 V_{DD}/2rB_0 R_{weight})*m_1 + (\mu_0 V_{bias}/2rB_0 R_{bias}) \quad (9)$$

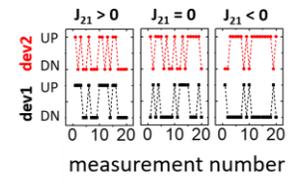
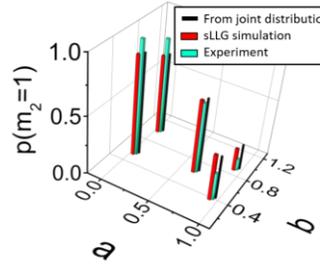
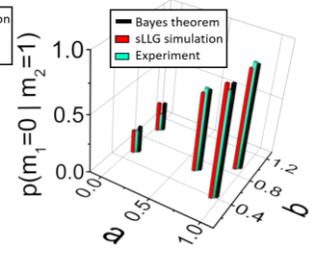

Fig. 3. Testing of the two node BN circuit. (a) Five different combinations of the CPT parameters that are experimentally implemented in hardware. (b) Representative sections of the measured data for positive, negative and no connection between device 1 and device 2 as shown in Fig 2(c). (c) Obtained output probabilities of cheese being stale for the five different given CPTs. The experimentally obtained probability values are in good agreement with theory and stochastic LLG simulations. (d) Inference about probability of the packaging being bad quality given that a stale cheese is found is plotted for the different CPTs, showing good match between direct experimental observation, Bayes theorem and stochastic LLG simulations.

For $R_{weight} = \infty$, which means $J_{21}=0$, the coefficient of $m_1$ in eq. 9 vanishes, and so does the correlation between the two devices. For a finite $R_{weight}$, the connection polarity dictates the sign of the correlation between the two devices, with a strength inversely proportional to $R_{weight}$. $V_{bias}$ makes the correlation asymmetric as its corresponding term in eqn. 9 does not change sign with the state of $m_1$. Therefore, we can span all possible conditional probabilities between two nodes of a BN (given by 'a' and 'b' in the CPT) by changing the circuit parameters $R_{weight}$, polarity and $R_{bias}$.

### C. Experimental testing of the hardware Bayesian network

We take five different CPTs with "a" and "b" spanning the range between 0 and 1, shown in Fig. 3 (a). We then calculate $J_{21}$ and $h_2$ for these five cases and design our circuit according to eq. (7) and eq. (8). The designed circuits are then tested by repeating a sequential pulsing scheme. The inset of Fig. 2 (c) shows the timing diagram of the measurement procedure. The two devices are pulsed sequentially by a Keithley 6221 current source that provides the clocking scheme for our devices. During the pulsing of the second device, a constant DC read current is passed through the first device in order to generate the input voltage to the second device. Then, this sequential pulsing is repeated to generate the required statistics. The two devices produce random outputs, but with correlated statistics, as is required by the CPT between the two random variables.

The output after each pulse is measured by a lock-in amplifier and then digitized. Representative sections of the device outputs are shown in Fig. 3 (b) for three different connection configuration. It is worth noting that the pulsing method being followed in the presented experiments (shown in the inset of Fig. 2(c)) is analogous to Gibbs sampling[26,27], which is widely used is used for statistical inference[28,29]. Here, each node of the network is pulsed (sampled) sequentially under the influence of all the other nodes, which are fixed to their current values.

The probability of finding a stale cheese block can be found from the joint probability distribution by using the probability chain rule:

$$p(m_2=1) = \Sigma_{m_1} p(m_1, m_2=1) = \Sigma_{m_1} p(m_2=1|m_1) \times p(m_1)$$
$$= p(m_2=1|m_1=0) \times p(m_1=0) + p(m_2=1|m_1=1) \times p(m_1=1)$$
$$= a \times p(m_1=0) + b \times p(m_1=1) \quad (10)$$

where $p(m_1 = 0$ or $1)$ is an input parameter. The number of terms in the above expression grows as $2^N$ where N is the number of parent nodes for the particular child node of interest[3]. Instead of performing this algebra, the required probability can be obtained from the circuit by directly observing the stochastic output of device 2 and obtaining its mean value over several pulsing cycles. This luxury of having to observe only the nodes of interest while disregarding all the other nodes is an advantage of using a probabilistic algorithm, versus calculating the probabilities using deterministic rules as discussed by Feynman[30] and utilized in many sampling schemes[26].

Similarly, given that a randomly drawn cheese block from a large lot is stale, the probability that it was caused by a low quality packaging material can be found by using Bayes theorem:

$$p(m_1=0|m_2=1) = p(m_1=0, m_2=1)/p(m_2=1)$$
$$= p(m_2=1|m_1=0) \times p(m_1=0)/p(m_2=1)$$
$$= [a \times p(m_1=0)] / [a \times p(m_1=0) + b \times p(m_1=1)] \quad (11)$$

The number of terms required in the evaluation of the above expression also grows as $\sim 2^N$ where "N" is the number of potential binary causes of a particular effect[3]. However, from the hardware BN, this probability can be directly obtained by observing the joint distribution of states of the two devices. It is to be noted here that this way of performing the inference always involves observing the joint distributions of only two nodes of the BN: nodes corresponding to the effect and the potential cause of interest, irrespective of N.

In our experiment, after 100 pulsing cycles, the obtained output probabilities for all the five circuits (representing the five different CPTs of Fig. 3(a)) is comparable with the expectation from calculating the joint probability distribution and is also verified by stochastic LLG simulations, as shown in Fig. 3 (c). Similarly, the obtained probabilities from inference is comparable with that from Bayes theorem and stochastic LLG simulations, seen in Fig. 3 (d).

### III. SIMULATION OF A FOUR NODE BAYESIAN NETWORK

In this section, we present a self-consistently coupled sLLG simulation of the more complicated, four node Bayesian network shown in the top left inset of Fig. 4 (a). Here, the BN consists of four nodes: cloud (C), rain (R), sprinkler (S), and wetness of grass (W). In this case, the evaluation of a node probability from the joint probability distribution requires the following evaluation, for example for the W node:

$$P(W) = \Sigma_C \Sigma_R \Sigma_S P(C, R, S, W) =$$
$$\Sigma_C \Sigma_R \Sigma_S P(C) P(R|C) P(S|C) P(W|RS) \quad (12)$$

Here the number of terms to be evaluated in the summation is eight, as each of the C, R and S nodes could take two possible values "0" or "1". Similarly performing inference, for example, what is the probability that it had rained, given that the grass is wet requires the following evaluation:

$$P(R|W) = P(R, W)/P(W) \quad (13)$$

where both the numerator and the denominator of the right-hand side of the above equation must be evaluated by summing over the joint probability distribution $P(C, R, S, W)$, resulting in the evaluation of four and eight terms respectively. However, by using the hardware, the required node probabilities and the inference can be obtained in exactly the same way as our previous two-node example: we simply observe the stochastic output of the corresponding node for probability assessment; and observe the joint distribution of only the R and the W node to perform the required inference. This is demonstrated in the simulation study below.

The parameters used in the sLLG simulation platform such as the magnet dimensions and the output sigmoidal response are benchmarked with the experimental results from the device in Fig. 1 (c). The coupling and biases are benchmarked with the two node BN network experiments shown in Fig. 2 and 3.

Fig. 4 (a) shows the circuit implementation, where each node is represented by a hardware p-bit as described in Fig. 1. It is to be noted here that an auxiliary p-bit (represented by node 'X') is needed to implement this four node Bayesian network. This is because, the CPT capturing the dependency of node 'W' on node 'R' and 'S' has four conditional probabilities, which can take any value between 0 and 1 independent of each other. Therefore, from basic principles of linear algebra, we need four independent physical parameters to implement this CPT. Two of the four required parameters are provided by the two interconnection weights ($J_{WR}$ and $J_{WS}$) and another parameter is provided by the bias to the node 'W' ($h_W$). The remaining one parameter is provided by the interconnection to the auxiliary node 'X'. The requirement of auxiliary nodes in designing Bayesian networks from p-bits is described in more detail by Faria et al.[18]. The dynamics of the PMA magnet used in the hardware p-bit design is captured by solving the sLLG equation with a monodomain macrospin assumption:

$$(1 + \alpha^2) \frac{d\hat{m}}{dt} = -|\gamma|\hat{m} \times \vec{H} - \alpha|\gamma|\hat{m} \times \hat{m} \times \vec{H} - \frac{1}{qN_s}\hat{m} \times \hat{m} \times \vec{I_s} + \frac{\alpha}{qN_s}\hat{m} \times \vec{I_s} \quad (14)$$

where, $\vec{H}$ is the total internal and external field along with thermal noise field, $\vec{I_s}$ is the spin current, $N_s = M_s V$ is the total magnetic moment with $M_s$ being the saturation magnetization, $\alpha$ is the damping coefficient, $\gamma$ is the gyromagnetic ratio. Magnet parameters used in the simulation are: $H_k = 200$ Oe,

$M_s = 1000$ emu/cc, $D_1 = 1\ \mu m$, $D_1 = 3\ \mu m$, $t = 1\ nm$, $\alpha = 0.1$. The average magnetization of each p-bit can be approximated by $\langle m_z \rangle = tanh\left(\frac{H}{H_0}\right)$, where $H$ is the Oersted field generated from the current coil and $H_0$ is a fitting parameter.

The coupling and bias component of $H_i$ can be realized through the coupling resistance $R_{weight}$ and $R_{bias}$ respectively with a mapping principle as described in eq. (7) and (8) for the two node case.

While solving the coupled sLLG, each p-bit is put along the hard axis by the GSHE current in a sequential order from

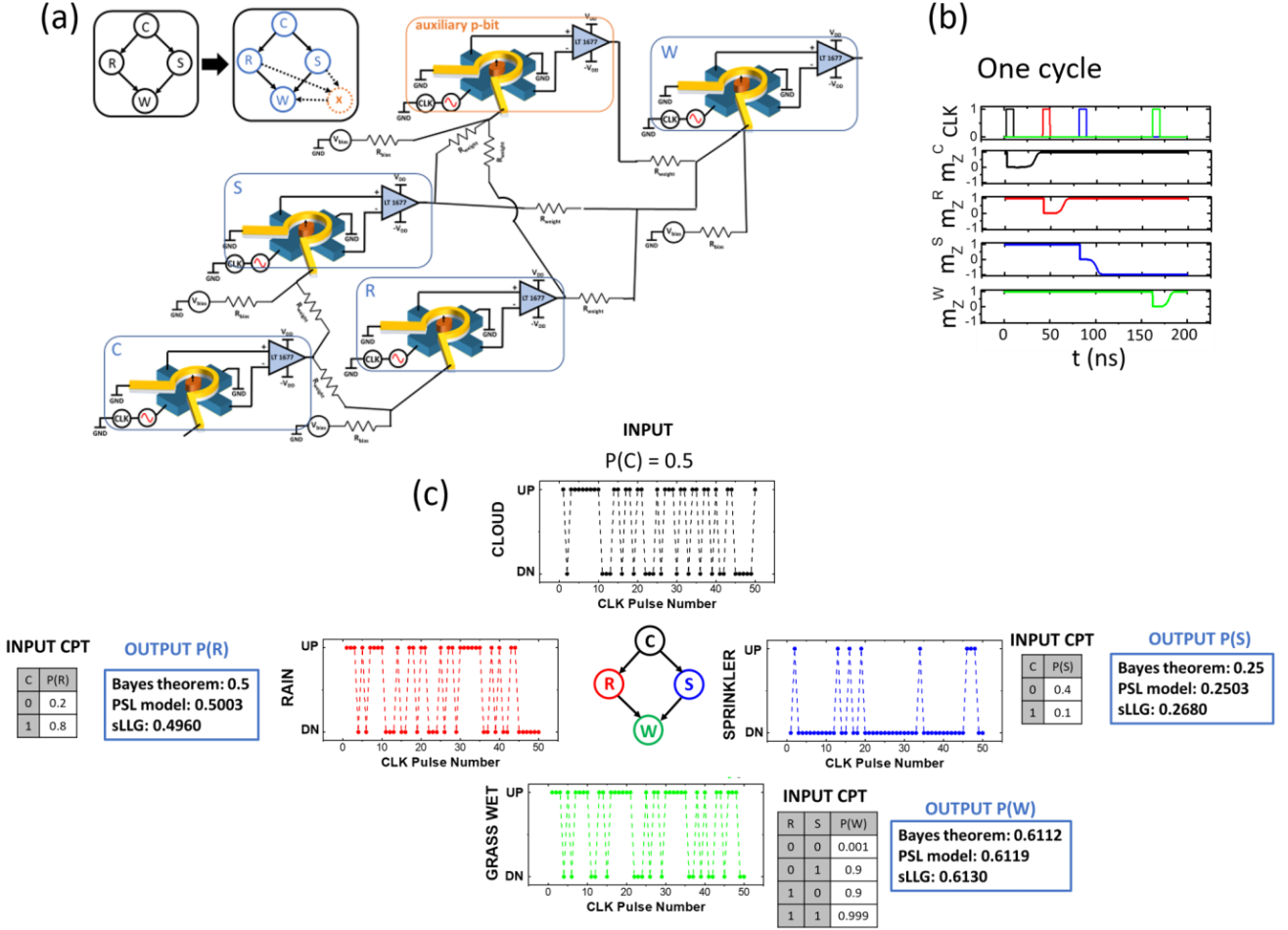

Fig. 4. Simulation results of a four node BN. (a) Hardware implementation layout (b) Representative one clock cycle of operation (c) Results obtained from the four node BN with the given CPTs shown as the input tables. Probabilities for each node, generated after 500 clock cycles are shown inside the blue boxes. Representative sections of the state of each node after 50 pulses is shown next to them. The obtained probabilities show good agreement with expectation from calculating the joint probability distribution.

For the system simulation, we start with chosen CPTs for each of the nodes. These are shown as the inputs next to the respective nodes in Fig. 4 (c). These values are then translated into coupling term $J_{ij}$ and bias term $h_i$ by following similar principles as in deriving eq. (5) and (6). The derivation for $J_{ij}$ and $h_i$ for an n-node Bayesian network is provided by Faria et al.[18]. The dimensionless terms $J_{ij}$ and $h_i$ are then translated to corresponding Oersted fields to each p-bit by a relation:

$$H_i = H_0 \sum_j J_{ij} m_j + h_i \qquad (15)$$

parent to child node and the magnetizations of all p-bits are recorded after their corresponding pulse is turned off. It is worth noting that the pulse sequence is important for the proper operation of the Bayesian network. The pulsing should start from the first node and move down the hierarchy from parent to corresponding child nodes. The order of pulsing among different nodes on the same hierarchy level (e.g. node R and S in our example) is not critical. Taking these principles into account, the pulsing order for one cycle is shown in Fig. 4 (b). This cycle is repeated several times to generate the probabilities of each of the four nodes. Fig. 4 (c) shows representative data of magnetization of each node for 50 pulses. From this distribution of the magnetization state of each node in 'UP' vs. 'DN' state, probabilities of each node are calculated. For example, the magnetization of the p-bit corresponding to 'sprinkler' node shows more occurrences in the 'DN' state compared to 'UP' state, resulting in a low probability of sprinkler being ON (P(S) ~ 0.25 in this case). Similarly, the probability of 'rain': P(R) and the probability of 'grass being wet': P(W) are obtained from the magnetization state distribution. The obtained probabilities are compared with those obtained by calculating the joint probability distribution as shown in the output tables alongside each of the four nodes in Fig. 4 (c). It can be seen that the probabilities obtained from the coupled sLLG result match well with the simple PSL behavioral model and with the values obtained from the evaluation of equation (12). Similarly, the probability of rain, given that the grass is wet ($P(R|W)$) is obtained from the coupled sLLG result is 0.73, which is well matched with the value of 0.75 obtained from equation (13). It is to be noted that the accuracy in this depends on the number of samples taken to calculate the probabilities.

## IV. CIRCUIT IMPLICATIONS AND IMPROVEMENTS

Previously proposed hardware implementations of Bayesian networks have used CMOS based pseudo random number generators realized with XOR-SHIFT circuits[6] or RAM-based Linear Feedback Gaussian Random Number Generators[4,5] that require a large area footprint. What we have demonstrated here is a compact true random number generator (TRNG) capable of operating at few hundreds of MHz. The speed of the device is only limited by the time required for SOT hard axis initialization and magnetization relaxation after removal of SOT, which normally requires a few nanoseconds as shown in the sLLG time plot panels of Fig. 1 (b). Compared to previously demonstrated spin based TRNG[33–35], this implementation employs a different scheme to generate random numbers. In our approach, any applied current that is larger than that required for hard axis initialization of the magnet will result in the generation of a random number with the correct statistic once the current pulse is removed[17,22]. Hence, in a large network, the device to device variation in the required current can be easily mitigated by choosing the largest value of the required current among all devices. Possible variations in the shape and offset of the sigmoidal activation function of our devices can be controlled by appropriately choosing the parameters $B_0$ and $h_2$ while translating the given CPTs into the connection weights, shown in equations (5)-(8). Also note that the Bayesian network proposed here does not require analog voltage sources or CMOS MUX to realize the CPT as proposed previously by Shim et al[22]. Using current controlled tunability of the device and auxiliary nodes, any CPT can be realized by using only p-bits, one voltage level ($V_{DD}$) and analog memristive elements for interconnections and individual biases similar to RRAM based neural networks. Such programmable analog memristive elements have been successfully demonstrated recently[36,37]. The energy requirement of the device demonstrated here can be improved by using the voltage-controlled magnetism (VCM) effect for hard-axis initialization as proposed by Scott et al.[38] in their benchmarking study (section IV of the main text). In addition, employing magnetic tunnel junctions (MTJs) instead of AHE can eliminate the need for OP-AMPs for readout. The typical difference in the two stable resistive states of an MTJ is of the order of 10 KΩ, whereas in case of AHE, it is a few ohms for standard material stacks. This allows the elimination of the OP-AMPs for readout. Implementations of an MTJ based readout scheme, where the state of the free layer magnet is converted to a voltage by a potential divider formed by the MTJ and a normal resistor was presented by Camsari et al.[12] (figure 3 of the main text) and Hassan et al.[39] (figure 4 of the main text). In these references, the voltage swing generated at the output is large enough to be converted to a "rail-to-rail" swing by a single inverter. In the above references, the MTJ free layer was designed to be a low barrier magnet, but the analysis of the output swing remains unchanged for our hard axis initialization scheme with stable magnets.

## V. CONCLUSION

We have experimentally demonstrated that by connecting two stochastic spintronic devices and designing the connection and bias parameters, BN building blocks can be implemented in hardware. By implementing BNs using such hardware, both probability assessment and inference can be performed by sampling the output of only the relevant nodes. Using experimentally benchmarked sLLG simulations, we have shown that a four node BN implemented in hardware using the presented stochastic devices can generate probabilities that are well matched to the theoretical values from calculating the joint probability distribution. This demonstration serves as a step towards building large scale hardware systems for implementing Bayesian networks.

### ACKNOWLEDGEMENT

This work was supported by the Center for Probabilistic Spin Logic for Low-Energy Boolean and Non-Boolean Computing (CAPSL), one of the Nanoelectronic ComputingResearch (nCORE) Centers as task 2759.003 and 2759.004, a Semiconductor Research Corporation (SRC) program sponsored by the NSF through CCF 1739635.

### AUTHOR CONTRIBUTIONS

P.D. performed the experiments with V.O.'s help. R.F. performed the numerical simulations. S.D., J.A. and Z.C. supervised the project and helped in the analysis of the results. P.D. wrote the manuscript with inputs from all authors. All authors reviewed the manuscript.

### ADDITIONAL INFORMATION

**Competing Interests:** The authors declare no competing interests.